\documentclass[]{raa}            
\usepackage{graphicx,times}
\usepackage{natbib}

\providecommand{\bm}[1]{\mbox{\boldmath$#1$\unboldmath}}

\begin{document}

   \title{A likely candidate of type Ia supernova progenitors:  the X-ray pulsating companion of the hot subdwarf HD 49798$^*$
   \footnotetext{\small $*$ Supported by the National Natural
Science Foundation of China.} }

 \volnopage{ {\bf 2010} Vol.\ {\bf X} No. {\bf XX}, 000--000}
   \setcounter{page}{1}

   \author{Bo Wang
      \inst{1,2,3}
   \and Zhan-Wen Han
      \inst{1,2}
   }

   \institute{National Astronomical Observatories/Yunnan
Observatory, Chinese Academy of Sciences, Kunming 650011, China; {\it wangbo@ynao.ac.cn, zhanwenhan@ynao.ac.cn}\\
\and
   Key Laboratory for the Structure and Evolution of Celestial
Objects, Chinese Academy of Sciences, Kunming 650011, China\\
        \and
             Graduate University of Chinese Academy of Sciences, Beijing
100049, China}

\abstract{HD 49798 is a hydrogen depleted subdwarf O6 star and has
an X-ray pulsating companion (RX J0648.0$-$4418). The X-ray
pulsating companion is a massive white dwarf. Employing Eggleton's
stellar evolution code with the optically thick wind assumption, we
find that the hot subdwarf HD 49798 and its X-ray pulsating
companion could produce a type Ia supernova (SN Ia) in future
evolution. This implies that the binary system is a likely candidate
of SN Ia progenitors. We also discussed the possibilities of some
other WD + He star systems (e.g. V445 Pup and KPD 1930+2752) for
producing SNe Ia.  }

   \authorrunning{B. Wang \& Z.-W. Han}            
   \titlerunning{A likely candidate of SN Ia progenitors: the companion of HD 49798}  
   \maketitle


%
%
\section{Introduction}           
\label{sect:intro}

Type Ia supernova (SN Ia) explosions are among the most energetic
events observed in the Universe. They appear to be good cosmological
distance indicators owing to their high luminosities and remarkable
uniformity, and have been applied successfully in determining
cosmological parameters (e.g. \textbf{$\Omega_{M}$} and
\textbf{$\Omega_{\Lambda}$}; Riess et al. 1998; Perlmutter et al.
1999). However, the exact explosion mechanism and the nature of
progenitors are still poorly understood (Hillebrandt \& Niemeyer
2000; Podsiadlowski 2010; Wang et al. 2008a, 2010), and no SN Ia
progenitor system before the explosion has been conclusively
identified (Wang \& Han 2009, 2010a; Meng \& Yang 2010a).

It is widely accepted that SNe Ia arise from thermonuclear
explosions of carbon--oxygen white dwarfs (CO WDs) in binaries
(Nomoto et al. 1997; Livio 2000). Over the past few decades, two
groups of SN Ia progenitor models have been proposed, i.e. the
double-degenerate (DD) and single-degenerate (SD) models. The DD
model involves the merger of two CO WDs (Tutukov \& Yungelson 1981;
Iben \& Tutukov 1984; Webbink 1984; Han 1998). Although the DD model
might be able to account for the explosion of a few overluminous SNe
Ia (Howell et al. 2006; Gilfanov \& Bogd$\acute{\rm a}$n 2010), it
is still suffering from the theoretical difficulty that the mergers
of two WDs may lead to an accretion-induced collapse rather than
thermonuclear explosion (Nomoto \& Iben 1985; Saio \& Nomoto 1985;
Timmes et al. 1994). For the SD model, the companion could be a
main-sequence (MS) star or a slightly evolved star (WD + MS
channel), or a red-giant star (WD + RG channel) (Hachisu et al.
1996; Li $\&$ van den Heuvel 1997; Langer et al. 2000; Fedorova et
al. 2004; Han $\&$ Podsiadlowski 2004, 2006; Chen $\&$ Li 2007;
Ruiter et al. 2009; L\"{u} et al. 2009; Meng \& Yang 2010b; Wang, Li
\& Han 2010; Wang \& Han 2010b). Observationally, there is
increasing evidence indicating that at least some SNe Ia may come
from the SD model (Hansen 2003; Ruiz-Lapuente et al. 2004; Voss \&
Nelemans 2008; Wang et al. 2008b; Justham et al. 2009). Moreover,
the detections of variable Na I D lines (Patat et al. 2007; Blondin
et al. 2009; Simon et al. 2009) and derivation of smaller absorption
ratio $R_{\rm V}$ that is characteristic of circumstellar material
(CSM) dust (Wang et al. 2009c), perhaps suggests the presence of CSM
around a subclass of SNe Ia.

Recently, Wang et al. (2009a) studied the WD + He star channel of
the SD model to produce SNe Ia, in which a CO WD accretes material
from an He MS star or a slightly evolved He star to increase its
mass to the Chandrasekhar (Ch) mass. The study derived the parameter
spaces for the progenitors of SNe Ia. By using a detailed binary
population synthesis approach, Wang et al. (2009b) found that the
Galactic SN Ia birthrate from this channel is $\sim$$0.3\times
10^{-3}\ {\rm yr}^{-1}$, and that this channel may account for SNe
Ia with short delay times ($\sim$45$-$140\,Myr) from the star
formation to SN explosion (see also Wang \& Han 2010c).

Hot subdwarf stars are near the blue end of the horizontal branch in
the Hertzsprung-Russell diagram. A particularly interesting member
of this class is a subdwarf O star, HD 49798, which is one of the
brightest subdwarfs and also a single-lined spectroscopic binary
with an orbital period of 1.548\,d (Thackeray 1970; Stickland \&
Lloyd 1994). This hydrogen-deficient subdwarf of spectral type O6
has been studied extensively (Kudritzki \& Simon 1978; Hamann et al.
1981). Bisscheroux et al. (1997) showed that the hot subdwarf star
HD 49798 must have a degenerate C-O core, and is in the He-shell
burning phase, which can explain its high luminosity. Israel et al.
(1997) reported the detection of a pulsating soft X-ray source (RX
J0648.0$-$4418) with a pulsation period of $\sim$13\,s from this
binary. The X-ray spectrum of the source is very soft, but has a
high-energy excess. The discovery of the regular X-ray pulsations
must arise from a compact companion of either a neutron star or a WD
(e.g. Israel et al. 1997; Bisscheroux et al. 1997). Bisscheroux et
al. (1997) excluded a neutron star as the companion of HD 49798, and
showed that all observations are consistent with a weakly magnetized
massive WD, which is accreting material from the wind of its
subdwarf companion.


Kudritzki \& Simon (1978) estimated a mass of 0.7$-$2.7$\,M_{\odot}$
for the hot subdwarf HD 49798. This system is consistent with a
double spectroscopic binary, favored by the detection of a fast
X-ray pulsar source, for which all the orbital parameters (including
the masses of the two components) may be derived. With this purpose,
Mereghetti et al. (2009) recently observed HD 49798/RX
J0648.0$-$4418 in may 2008 with XMM-Newton satellite. They confirmed
that the 13\,s pulsation in the X-ray binary
HD49798/RXJ0648.0$-$4418 is due to a rapidly rotating WD. From the
pulse time delays and the system's inclination, constrained by the
duration of the X-ray eclipse discovered in this observation, they
derived the masses of the two components. The corresponding masses
are 1.50$\pm$0.05$\,M_{\odot}$ for HD 49798 and
1.28$\pm$0.05$\,M_{\odot}$ for the WD.

The existence of WD + He star systems is supported by some
observations (Wang et al. 2009a). The hot subdwarf HD 49798 with its
WD companion is such a system, and may be a candidate of SN Ia
progenitors. The goal of this paper is to investigate the evolution
and final fate of the hot subdwarf HD 49798 and its WD companion,
and to explore whether this binary system could produce an SN Ia. In
Section 2, we describe the numerical code of the binary evolution
and the input physics. In Section 3, we give the binary evolutionary
results. Finally, discussion and conclusion are given in Section 4.

\section{BINARY EVOLUTION CALCULATIONS}
We use Eggleton's stellar evolution code (Eggleton 1971, 1972, 1973)
to calculate the evolution of the hot subdwarf HD 49798 and its WD
companion. The code has been updated with the latest input physics
over the last four decades (Han et al. 1994; Pols et al. 1995,
1998). Roche lobe overflow (RLOF) is treated within the code as
described by Han et al. (2000). We set the ratio of mixing length to
local pressure scale height, $\alpha=l/H_{\rm p}$, to be 2.0. The
opacity tables are compiled by Guo et al. (2008). To simplify our
calculations, the He star was assumed to have a He abundance
$Y=0.98$ and metallicity $Z=0.02$. We assume that the binary model
starts with a 1.5\,$M_{\odot}$ He star and a 1.28$\,M_{\odot}$ CO WD
having a 1.548\,d orbit period, similar to the initial model of HD
49798 and its WD companion.

Instead of solving stellar structure equations of the WD, we use the
optically thick wind model (Hachisu et al. 1996) and adopt the
prescription of Kato \& Hachisu (2004, KH04) for the
mass-accumulation efficiency of He-shell flashes onto the WD. If the
mass-transfer rate, $|\dot M_2|$, is above a critical rate, $\dot
M_{\rm cr}$, we assume that He burns steadily on the surface of the
WD and that the He-rich matter is converted into C and O at a rate
$\dot M_{\rm cr}$. The unburned matter is lost from the system,
presumably in the form of the optically thick wind at a mass-loss
rate $\dot M_{\rm wind}=|\dot M_2| - \dot M_{\rm cr}$. The critical
mass-transfer rate is
\begin{equation}
\dot M_{\rm cr}=7.2\times 10^{-6}\,(M_{\rm
WD}/M_{\odot}-0.6)\,M_{\odot}\,\rm yr^{-1},
\end{equation}
based on WD models computed with constant mass-accretion rates
(Nomoto 1982). Similar to the work of Wang et al. (2009a), following
assumptions are adopted when $|\dot M_2|$ is smaller than $\dot
M_{\rm cr}$ (see also Wang \& Han 2010d). (1) If $|\dot M_2|$ is
less than $\dot M_{\rm cr}$ but higher than the minimum accretion
rate of stable He-shell burning, $\dot M_{\rm st}$ (KH04), it is
assumed that the He-shell burning is stable and that there is no
mass loss. (2) If $|\dot M_2|$ is less than $\dot M_{\rm st}$ but
higher than the minimum accretion rate of weak He-shell flashes,
$\dot M_{\rm low}=4.0\times10^{-8}\,M_{\odot}\,\rm yr^{-1}$ (Woosley
et al. 1986), He-shell flashes occur and a part of the envelope mass
is assumed to be blown off from the surface of the WD. The WD
mass-growth rate in this case is linearly interpolated from a grid
computed by KH04, where a wide range of WD masses and accretion
rates was calculated in the He-shell flashes. (3) If $|\dot M_2|$ is
lower than $\dot M_{\rm low}$, the He-shell flashes will be so
strong that no mass can be accumulated onto the WD.

We define the mass-growth rate of the CO WD, $\dot{M}_{\rm CO}$, as
 \begin{equation}
 \dot{M}_{\rm CO}=\eta _{\rm He}|\dot{M}_{\rm 2}|,
  \end{equation}
where $\eta _{\rm He}$ is the mass-accumulation efficiency for the
He-shell burning. According to the assumptions above, the values of
$\eta _{\rm He}$ are:
\begin{equation}
\eta_{\rm He}= \left\{ \begin{array}{l@{\quad,\quad}l}
\dot M_{\rm cr}\over |\dot M_2| &  |\dot{M_2}|>\dot{M}_{\rm cr},\strut\\
1\, &  \dot{M}_{\rm cr}\ge |\dot{M_2}|\ge \dot{M}_{\rm st},\strut\\
\eta'_{\rm He}\, &  \dot{M}_{\rm st}> |\dot{M_2}|\ge \dot{M}_{\rm low},\strut\\
0 &  |\dot{M_2}|< \dot{M}_{\rm low}.\strut\\
\end{array} \right.
\end{equation}

We incorporated the prescriptions above into Eggleton's stellar
evolution code and followed the evolution of the WD + He star
system. The mass lost from the system via the optically thick wind
is assumed to take away specific orbital angular momentum of the
He-accreting WD. We also consider the mass-loss from the stellar
wind of the He star, which is assumed to take away specific orbital
angular momentum of the He star. This mass-loss of the He star from
the stellar wind is considered as
\begin{equation}
\log\dot{M}_{\rm wind}=1.5\,\log L/L_{\odot} - 14.4,
\end{equation}
based on the stellar wind analysis from subdwarf O stars (including
He stars) to massive O stars (Jeffery \& Hamann 2010). We assume
that, if the WD grows to 1.4\,$M_{\odot}$, it explodes as an SN Ia.
Finally, we have calculated the evolution of the WD + He star
system, and found that this binary system could produce an SN Ia.

\begin{figure}
   \begin{center}
\includegraphics[width=6.8cm,angle=270]{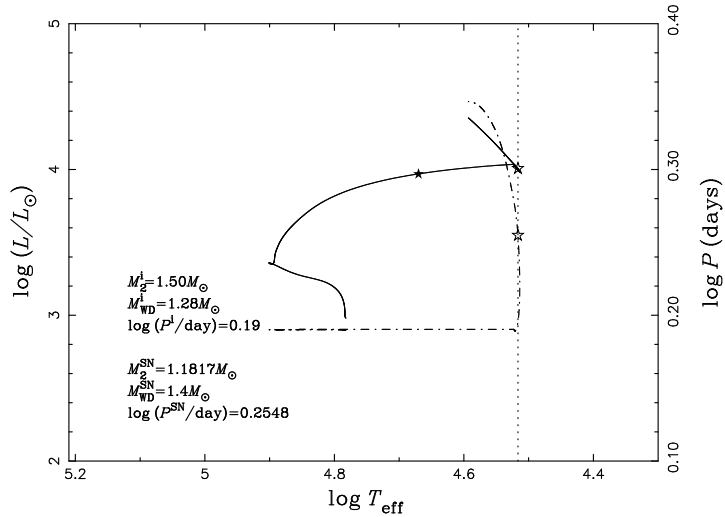}
 \caption{Results of binary evolution calculations with initial masses of two components and
the orbital period as well similar to the binary system HD49798/RX
J0648.9-4418. The evolutionary track of the He donor star is shown
as a solid curve, and the evolution of the orbital period is shown
as a dash-dotted curve. The solid star represents the current
position of HD 49798. Dotted vertical line and open stars indicate
the position where the WD is expected to explode as an SN Ia. The
initial binary parameters and the parameters at the moment of the SN
explosion are given in this figure.}
   \end{center}
\end{figure}

\section{BINARY EVOLUTION RESULTS} \label{3. BINARY EVOLUTION RESULTS}

The WD + He star system starts with ($M_2^{\rm i}$, $M_{\rm WD}^{\rm
i}$, $\log (P^{\rm i}/{\rm day})$) $=$ (1.50, 1.28, 0.19), where
$M_2^{\rm i}$, $M_{\rm WD}^{\rm i}$ are the initial masses of the He
star and of the CO WD in solar mass, and the $P^{\rm i}$ is the
initial orbital period in days. Figure 1 shows the evolutionary
track of the He donor star and the evolution of the orbital period,
where the current position of HD 49798 is also indicated. Figure 2
displays $\dot M_2$, $\dot M_{\rm CO}$ and $M_{\rm WD}$ varying with
time after HD 49798 fills its Roche lobe.

The He star undergoes the He-core burning for about 7\,million
years. After the exhaustion of the central helium, the envelope of
the He star expands. When the radius of the He star reaches
1.45$\,R_{\odot}$ (Bisscheroux et al. 1997), the He star evolves to
the current position of HD 49798. At this point, the CO core mass of
the He star reaches 0.79$\,M_{\odot}$, the logarithmic temperature
and luminosity are 4.67 and 3.97, respectively, which are consistent
with the parameters of HD 49798 derived from the detailed spectral
analysis (see Bisscheroux et al. 1997). After about
$4\times10^{4}$\,yr, HD 49798 begins to fill its Roche lobe due to
the rapid expansion of its envelope. The mass-transfer rate
$|\dot{M}_{\rm 2}|$ exceeds $\dot M_{\rm cr}$ soon after the onset
of RLOF, leading to a wind phase. In this phase, a part of the
transferred mass is blown off in the form of the optically thick
wind, and the left is accumulated onto the WD. The WD increases its
mass to 1.4\,$M_{\odot}$ after about $2\times10^{4}$\,yr and
explodes as an SN Ia. At this moment, the mass of the He star is
$M^{\rm SN}_2=1.1817\,M_{\odot}$ and the orbital period is $\log
(P^{\rm SN}/{\rm day})=0.2548$.

\begin{figure}
   \begin{center}
\includegraphics[width=6.8cm,angle=270]{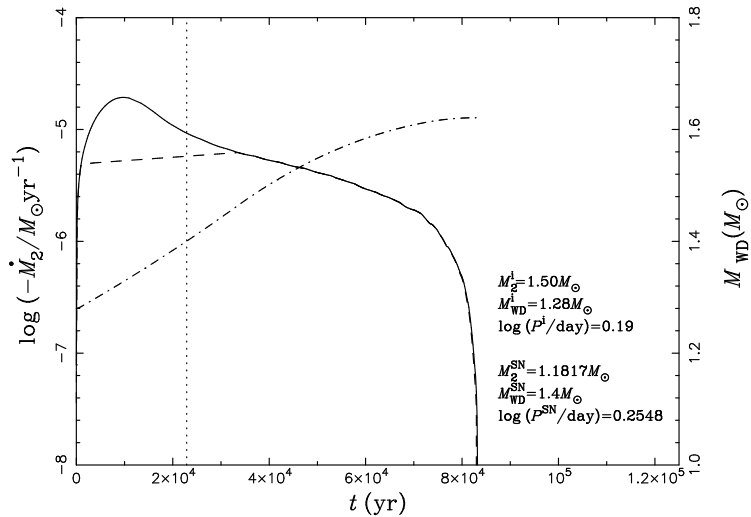}
 \caption{Results of binary evolution calculations, but for $\dot M_2$ (solid curve), $\dot
M_{\rm CO}$ (dashed curve) and $M_{\rm WD}$ (dash-dotted curve)
varying with time after HD 49798 fills its Roche lobe. Dotted
vertical line indicates the position where the WD is expected to
explode as an SN Ia.}
   \end{center}
\end{figure}

\section{DISCUSSION AND CONCLUSION} \label{5:DISCUSSION AND CONCLUSIONS}

The observations indicate that the X-ray source (RX J0648.0-4418) is
a weekly magnetized massive WD which is accreting matter from the
wind of its subdwarf companion. The mass loss of HD 49798 from the
wind in our calculations is about $3\times10^{-9}\,M_{\odot}\,\rm
yr^{-1}$. By using the Bondi-Hoyle formalism as described by
Davidson \& Ostriker (1973), we can make an estimate of mass $\dot
M_{\rm acc}$ captured from the wind by the gravitational field of
the WD, and the luminosity $L_{\rm acc}$ converted by the potential
energy in the process of accretion. We find that a wind velocity
between 800$-$1350\,km\,s$^{-1}$ with $\dot M_{\rm
wind}=3\times10^{-9}\,M_{\odot}\,\rm yr^{-1}$ will result in an
accretion luminosity between $10^{30}-10^{31}$\,erg\,s$^{-1}$,
consistent with that of the observed X-ray luminosity
$\sim$$10^{31}$\,erg\,s$^{-1}$ in the 0.2$-$10\,keV energy range
(e.g. Mereghetti et al. 2009).

Mereghetti et al. (2009) confirmed that RX J0648.0$-$4418 is a
rapidly rotating WD. The maximum stable mass of a rotating WD may be
above the standard Chandrasekhar (Ch) mass (e.g. Uenishi et al.
2003; Yoon \& Langer 2005; Chen \& Li 2009), and the maximum
possible mass a CO WD can reach by mass accretion is about
2.0$\,M_{\odot}$ (see Yoon \& Langer 2005). According to our
calculations, the maximum mass that the WD RX J0648.0$-$4418 can
reach is about $1.62\,M_\odot$ (see Fig. 2), which is larger than
the standard Ch mass ($1.4\,M_\odot$) we set in this paper. Thus, we
point out that the WD companion of HD 49798 might evolve towards a
thermonuclear explosion of the super-Ch mass WD, producing an
overluminous SN Ia (e.g. Howell et al. 2006). We also note that, if
rotation is taken into account, He burning is much less violent than
that without rotating (see Yoon et al. 2004). This may significantly
increase the He-accretion efficiency (i.e. $\eta _{\rm He}$ in our
parametrization). Therefore, more He-rich matter can be converted
into C and O, increasing the chance for a WD to survive above the Ch
mass limit.

The process that leads to the formation of this peculiar system is
still poorly understood. It is suggested that HD 49798/RX
J0648.0-4418 corresponds to a previously unobserved evolutionary
stage of a massive binary system, after the common-envelope phase
and spiral-in (e.g. Israel et al. 1997; Bisscheroux et al. 1997). A
primordial binary system with a primary mass  $M_{\rm
1,i}\sim5.0-8.0\,M_\odot$ and a secondary mass $M_{\rm
2,i}\sim2.0-6.5\,M_\odot$ may produce a system like HD 49798 and its
WD companion (Wang et al. 2009b). We note that the primordial binary
system has a short delay time ($\sim$100\,Myr) from the star
formation to SN explosion. Thus, HD 49798 and its WD companion can
contribute to the young population of SNe Ia revealed by recent
observations (Mannucci et al. 2006; Aubourg et al. 2008). The young
population of SNe Ia may have an effect on models of galactic
chemical evolution, since they would return large amounts of iron to
the interstellar medium much earlier than previously thought.

The WDs usually have masses in a narrow range centered at about
0.6$\,M_\odot$ (Kepler et al. 2007). However, a few examples of WDs
with very high mass ($>$1.2$\,M_\odot$) have recently been reported
(Dahn et al. 2004; Vennes \& Kawka 2008). And the X-ray source RX
J0648.9-4418 may be such a massive WD. These massive WDs in binary
systems are good candidates for the formation of SNe Ia if the mass
transfer can occur, since a small amount of accreted mass could
drive them above the Ch mass limit.

Besides HD 49798 and its WD companion, there are also some other WD
+ He star systems, e.g. V445 Pup and KPD 1930+2752, which are also
good candidates of SN Ia progenitors.  V445 Pup is an He nova (Ashok
\& Banerjee 2003; Kato \& Hachisu 2003). Kato et al. (2008) recently
presented a free-free emission dominated light curve model of V445
Pup, based on the optically thick wind theory (Hachisu et al. 1996).
The light curve fitting showed that the mass of the WD is above
$1.35\,M_{\odot}$, and half of the accreted matter remains on the
WD, resulting in the mass increase of the WD. Thus, V445 Pup is
suggested to be one of the best candidate of SN Ia progenitors (Kato
et al. 2008; see also Woudt et al. 2009). However, the orbital
period of the binary system and the mass of the He star are still
uncertain so far. To clarify the above parameters, further
observations of V445 Pup are needed when the dense dust shell
disappears. KPD 1930+2752 is regarded as another candidate of SN Ia
progenitors, giving rise to SN Ia explosion in the form of merging
WDs (Maxted et al. 2000; Geier et al. 2007). Note that, the DD model
is not supported theoretically, as it may lead to an
accretion-induced collapse rather than to an SN Ia (Nomoto \& Iben
1985; Saio \& Nomoto 1985; Timmes et al. 1994). On the other hand,
KPD 1930+2752 may also produce an SN Ia through the SD model.
However, the mass of the He donor star in KPD 1930+2752 is limited
to the range between 0.45$\,M_{\odot}$ and 0.52$\,M_{\odot}$ (Geier
et al. 2007), which is below the minimum mass (0.95$\,M_{\odot}$)
for producing SNe Ia (Wang et al. 2009a). Thus, KPD 1930+2752 may
not be a good candidate to produce an SN Ia via the SD model.

In this paper, by using the optically thick wind model (Hachisu et
al. 1996) and adopting the prescription of KH04 for the mass
accumulation efficiency of the He-shell flashes onto the WD, we
showed that the hot subdwarf HD 49798 and its X-ray pulsating
companion could produce an SN Ia in future evolution. We also
discussed the possibilities of some other WD + He star systems for
producing SNe Ia. To further study the WD + He star channel of SNe
Ia, large samples of WD + He star systems are expected in future
observations.

\normalem
\begin{acknowledgements}
We thank the anonymous referee for valuable comments that helped us
to improve the paper. This work is supported by the National Natural
Science Foundation of China (Grant No. 10821061), the National Basic
Research Program of China (Grant No. 2007CB815406) and the Chinese
Academy of Sciences (Grant No. KJCX2-YW-T24).
\end{acknowledgements}

\label{lastpage}

\end{document}